\documentclass[aps,prd,preprint,floatfix]{revtex4}

\pdfoutput=1
\usepackage[latin1]{inputenc}
\usepackage[english]{babel}
\usepackage{latexsym}
\usepackage{graphicx}
\usepackage{amsmath}
\usepackage{eucal}
\usepackage{natbib}
\usepackage{upgreek}

\usepackage{hyperref}

\newcommand{\LorTrans}[2]{\Lambda^{#1}_{\phantom{#1}#2}}
\newcommand{\InvLorTrans}[2]{\Lambda_{#1}^{\phantom{#1}#2}}

\newcommand{\tetrad}[2]{h^{#1}_{\phantom{#1}#2}}
\newcommand{\Trivtetrad}[2]{e^{#1}_{\phantom{#1}#2}}
\newcommand{\Invtetrad}[2]{h_{#1}^{\phantom{#1}#2}}

\newcommand{\Bfield}[2]{B^{#1}_{\phantom{#1}#2}}

\newcommand{\LorentzConn}[3]{A^{#1}_{\phantom{#1}#2#3}}

\newcommand{\SpaceTimeConn}[3]{\Gamma^{#1}_{\phantom{#1}#2#3}}
\newcommand{\Torsion}[3]{T^{#1}_{\phantom{#1}#2#3}}

\newcommand{\FockIvanenko}[1]{\mathcal{D}_{#1}}

\begin{document}

\title{
Flat spin connections in the Teleparallel equivalent of General Relativity
}
\author{A. Zubiaga}
\email{asier.zubiaga@gmail.com}
\affiliation{Independent Scientist, St. Gallen (Switzerland)}

\begin{abstract}
This work generalizes the treatment of flat spin connections in the teleparallel equivalent of general relativity. 
It is shown that a general flat spin connection form a subspace in the affine space of spin connections which is dynamically decoupled from the tetrad and the matter fields. A translation in the affine subspace introduces a torsion term without changing the tetrad. Instead, the change in the torsion is related to the introduction of a global acceleration field term that introduces Lorentz inertial effects in the reference frame. The dynamics of the gravitationally coupled matter fields remains however equivalent regardless of the flat spin connection chosen. The implications of the break of this invariance by a general $f(T)$ and $f(R)$ is discussed.
\end{abstract}

\maketitle

%\tableofcontents

\section{Introduction}
The description of gravitational interaction changed after A. Einstein proposed his General Relativity (GR) theory~\cite{einstein_grundlage_1916} to extend the Special Relativity (SR) for accelerated observers and to include the gravitational interaction. Since then, predictions of GR such as the deflection of light~\cite{dyson_deflection_light_1920}, the precession of the perihelion of Mercury~\cite{einstein_grundlage_1916}, the gravitational redshift of light~\cite{pound_gravitational_1959}, or the gravitational waves~\cite{ligo_virgo_gwave_2016_short} have been tested to great detail. GR thus remains as the best description of gravitational phenomena. Still, important challenges remain mostly related to the quantum formulation of the theory. Reaching a formulation of a normalizable quantum theory of gravitation with predictive power will require deeper understanding of the nature of gravitational processes. The loop quantum gravity~\cite{ashtekar_background_2004} and spin foams~\cite{perez_spin-foam_2013} are promising candidates for a quantum theory of gravitation, and the string theory~\cite{green_superstring_2012} promises to unify the description of the four fundamental interactions. However, the research program followed by both approaches have evolved with difficulties and many open questions remain. 

The Teleparallel formulation of Gravity (TG) is a particular case of the Poincare Gauge Theories~\cite{RevModPhys.48.393} which implement the local Poincare symmetry in addition to the covariance under coordinate transformations in a metric gravitation theory. The teleparalell equivalent of GR (TEGR) uses flat spin connections $\LorentzConn{a}{b}{\mu}$, with vanishing curvature and non-zero torsion, to extend the invariance under global Lorentz transformations of SR to local Lorentz transformations in the internal algebraic space. The TEGR was formulated by Hayashi {\it et al}.~\cite{hayashi_new_1979}, and it was subsequently extensively discussed by Aldovandri {\it et al}.~\cite{aldrovandi_teleparallel}. The canonical formalism has been also investigated in Blagojevi\ifmmode \acute{c}\else \'{c}\fi{} {\it et al}.~\cite{blagojevic_hamiltonian_2000,blagojevic_gauge_2000,PRD64_Blagojevic}, Maluf {\it et al}.~\cite{jmc35_Maluf,maluf_hamiltonian_2001} and Ferraro {\it et al}.~\cite{ferraro_hamiltonian_2016}  

Contrary to the Levi-Civita connection of GR, $\LorentzConn{a}{b}{\mu}$ is not uniquely determined by the tetrad. Often the space-time counterpart, the Weitzenb\"ock connection $\Gamma^{\alpha}_{\phantom{\alpha}\beta\gamma} = \Invtetrad{a}{\rho} \partial_{\gamma}\tetrad{a}{\beta}$, is used with $\LorentzConn{a}{b}{\mu}$ set to zero. This work analyzes the role of non-zero $\LorentzConn{a}{b}{\mu}$ in the teleparallel formulation of gravity (TG). It discusses also the consequences for $f(T)$ and $f(R)$ modifications. Section~\ref{sec_tegr} describes the main characteristics of the TEGR spacetime, the Weitzenb\"ock spacetime, and section~\ref{sec_lagrangian} gives the Lagrangian formulation and the Euler-Lagrange equations. Section~\ref{sec_conn} introduces a gauge transformation of the flat spin connection that leaves invariant the dynamics of the gravitational system coupled to matter fields and discusses the kinematical character of the flat spin connection. Finally, section~\ref{sec_discuss} discusses the consequences of non-observable $\LorentzConn{a}{b}{\mu}$ for $f(T)$ and $f(R)$ modifications of TEGR. The main conclusions are given in section~\ref{sec_conc}. Throughout the text, latin $\{a,b,c,...\}$ and greek $\{\mu,\nu,\rho,...\}$ letters are reserved for indices of the internal (Minkowski) space and the Riemann spacetime, respectively. 

\section{Weitzenb\"ock spacetime}~\label{sec_tegr}
The teleparallel spacetime is composed by an internal algebraic tensor space with a Minkowski metric $\eta_{ab}$ and a Riemann spacetime with coordinate $\{x^{\mu}\}$.~\cite{aldrovandi_teleparallel} The fundamental quantities describing the gravitational field are the tetrad form $\tetrad{a}{\mu}(x)$ and a flat spin connection $A_{\mu} = \frac{1}{2}\LorentzConn{a}{b}{\mu} \hat{S}_{ab}$, a one-form which assumes values in the Lie algebra of the Lorentz group $ \hat{S}_{ab}$, and has null curvature
\begin{equation}
%R^a_{\phantom{a}b\mu\nu} = \partial_{\mu}A^{a}_{\phantom{a}b\nu} - \partial_{\nu}A^{a}_{\phantom{a}b\mu} + A^{a}_{\phantom{a}c\mu}A^{c}_{\phantom{b}b\nu} - A^{a}_{\phantom{a}c\nu}A^{c}_{\phantom{c}b\mu} = 0.
R^a_{\phantom{a}b\mu\nu} = 
\partial_{\mu}A^{a}_{\phantom{a}b\nu} + A^{a}_{\phantom{a}c\mu}A^{c}_{\phantom{b}b\nu} -  \left( \mu \leftrightarrow \nu \right) = 0.
\end{equation}
The torsion, on the other hand, is not zero 
 \begin{equation}
T^a_{\phantom{a}\mu\nu} = 
\partial_{\mu}\tetrad{a}{\nu} + \LorentzConn{a}{b}{\mu}\tetrad{b}{\nu} - \left( \mu \leftrightarrow \nu \right) \ne 0.
\end{equation}
The tangent bundle of the Riemann spactime and the internal space are soldered by the tetrad form $\tetrad{a}{\mu}(x)$. The inverse of the tetrad form $\Invtetrad{a}{\mu}(x)$ is determined by $\tetrad{a}{\mu}(x)$ $\Invtetrad{b}{\mu}(x)$ = $\delta^a_b$ and $\tetrad{a}{\mu}(x)$ $\Invtetrad{a}{\nu}(x)$ = $\delta^{\nu}_{\mu}$. The metric of the Riemann spacetime is 
\begin{equation}\label{eq_metric_rel}
g_{\mu\nu}= \eta_{ab} \tetrad{a}{\mu} \tetrad{b}{\nu}, 
\end{equation}
Using this relation the tetrad can be determined from the Riemann metric up to a local Lorentz transformation, $h'^a = \LorTrans{a}{b}(x_{\mu}) h^b$. Space-time forms $\phi_{\mu}$ and Minkowski forms $\phi_a$ are soldered by the tetrad $\phi_{\mu} = \tetrad{a}{\mu} \phi_a$, and space-time vectors $\phi^{\mu}$ and Minkowski vectors $\phi^a$ by the inverse tetrad $\Invtetrad{a}{\mu}$ $\phi^{\mu} = \Invtetrad{a}{\mu} \phi^a$. The Fock-Ivanenko covariant derivative~\cite{Fock_Ivanenko} $\mathcal{D}_{\mu}\phi^a$, is defined in the Teleparallel spacetime with the flat spin connection $\LorentzConn{a}{b}{\mu}$. For a vector $\phi^a$ it writes as
\begin{equation}
\mathcal{D}_{\mu}\phi^a = \partial_{\mu} \phi^a + A^{a}_{\phantom{a}b\mu} \phi^b. 
\end{equation}
The soldering condition $\Invtetrad{a}{\rho}\mathcal{D}_{\nu}\phi^a=\nabla_{\nu}\phi^{\rho}$ links the Fock-Ivanenko derivative to the spacetime covariant derivative 
\begin{equation}
\nabla_{\nu}\phi^{\rho} = \partial_{\nu}\phi^{\rho} + \Gamma^{\rho}_{\phantom{\rho}\mu\nu} \phi^{\mu}
\end{equation}
which uses the Weitzenb\"ock spacetime connection~\cite{de_andrade_teleparallel_2001}  with non-zero $\LorentzConn{a}{b}{\mu}$:
\begin{equation}
\Gamma^{\rho}_{\phantom{\rho}\mu\nu} = \Invtetrad{a}{\rho}\partial_{\nu} \tetrad{a}{\mu} +  \Invtetrad{a}{\rho}A^{a}_{\phantom{a}b\nu} \tetrad{b}{\mu}
\end{equation}
Both connections fulfill the metric compatibility condition. 
\begin{equation}\label{eq_tetrad_consist}
\partial_{\nu} \tetrad{a}{\mu} + \LorentzConn{a}{b}{\nu} \tetrad{b}{\mu} - 
\SpaceTimeConn{\beta}{\mu}{\nu} \tetrad{a}{\beta} = 0
\end{equation}
Under a local Lorentz transformation $\LorTrans{a}{b}(x)$ in the internal Minkowski space, the Fock-Ivanenko derivative transforms covariantly, 
\begin{eqnarray}
\mathcal{D}_{\mu}\phi'^a &=& \LorTrans{a}{b}(x) \mathcal{D}_{\mu}\phi^b
\end{eqnarray}
and the connection acquires a vacuum component
\begin{equation}
A'^{a}_{\phantom{a}b\mu} = 
\LorTrans{a}{c}(x) \InvLorTrans{b}{d}(x) A^{c}_{\phantom{c}d\mu} + \LorTrans{a}{c}(x) \partial_{\mu} \InvLorTrans{b}{c}(x) 
\end{equation}

Nonetheless, the equivalence principle requires that the minimal coupling of gravity to matter uses the Levi-Civita connection,  
\begin{equation}
\mathring{\omega}^a_{\phantom{a}b\mu} = \LorentzConn{a}{b}{\mu}-K^a_{\phantom{a}b\mu}
 = - \frac{1}{2}\left( t^{\phantom{b}a}_{b\phantom{a}\mu} + t^{\phantom{\mu}a}_{\mu\phantom{a}b} - t^{a}_{\phantom{a}b\mu} \right)
\end{equation}
with $K^a_{\phantom{a}b\mu} = 1/2\left( T^{\phantom{b}a}_{b\phantom{a}\mu} + T^{\phantom{\mu}a}_{\mu\phantom{a}b} - T^{a}_{\phantom{a}b\mu} \right)$  the contortion tensor of $T^a_{\phantom{a}b\mu}$ and $t^{a}_{\phantom{a}\mu\nu} = \partial_{\mu} \tetrad{a}{\nu} - \partial_{\nu} \tetrad{a}{\mu}$. Although $\LorentzConn{a}{b}{\mu}$ enters the definition of $\mathring{\omega}^a_{\phantom{a}b\mu}$ as the only connection of the theory, notably, it is missing from the final expression.

\section{Lagrangian formulation}~\label{sec_lagrangian}
The Lagrangian density of the TEGR is a written using $\tetrad{a}{\mu}$ and $\LorentzConn{a}{b}{\mu}$ and it includes kinetic terms for $\partial_{\nu}\tetrad{a}{\mu}$
\begin{equation}
\mathcal{L}_G = \frac{h}{2\kappa} \left(
\frac{1}{4} T^{\rho}_{\phantom{\rho}\mu\nu} T_{\rho}^{\phantom{\rho}\mu\nu}
+ \frac{1}{2} T^{\rho}_{\phantom{\rho}\mu\nu} T_{\phantom{\nu\mu}\rho}^{\nu\mu}
- T^{\rho}_{\phantom{\rho}\mu\rho} T_{\phantom{\nu\mu}\nu}^{\nu\mu}
\right)\label{eq_lagrangian}
\end{equation}
$\kappa = 8\pi G/c^4$ and $G$ the Newton's constant. It is equivalent to the Lagrangian density of GR up to a divergence surface terms~\cite{jmc35_Maluf} 
\begin{equation}\label{eq_lagrangian_diverg}
\mathcal{L}_G = \frac{1}{2\kappa} \sqrt{-g} \mathring{R} - \partial_{\mu}\left(\frac{h}{\kappa} T^{\nu\mu}_{\phantom{\nu\mu}\nu}\right)
\end{equation}
with $\mathring{R}$ the curvature of $\mathring{\omega}^a_{\phantom{a}b\mu}$. The counterpart of the equation of Einstein is obtained varying the total lagrangian against the tetrad $\delta\left(\mathcal{L}_G+\mathcal{L}_M\right)/\delta\tetrad{a}{\mu}=0$:
\begin{equation}\label{eq_grav_field}
%Gravitational field equation
\partial_{\sigma}\left(h S_a^{\phantom{a}\rho\sigma}\right) - kh \jmath_a^{\rho} = k\Theta_a^{\phantom{a}\rho}
\end{equation}
with $h \Theta_a^{\phantom{a}\rho} = - \delta \mathcal{L}_M / \delta h^a_{\phantom{a}\rho} $ the energy-momentum source tensor, $S_a^{\phantom{a}\rho\sigma}$ the superpotential
%Superpotential
\begin{equation}
S_a^{\phantom{a}\rho\sigma} = K^{\rho\sigma}_{\phantom{\rho\sigma}a} 
- h_a^{\sigma} T^{\theta\rho}_{\phantom{\theta\rho}\theta}
+ h_a^{\rho} T^{\theta\sigma}_{\phantom{\theta\sigma}\theta}
\end{equation} 
and $\jmath_a^{\phantom{a}\rho}$ the gauge current
%Gauge current
\begin{equation} 
\jmath_a^{\rho} = \frac{1}{k} h_a^{\phantom{a}\lambda} S_c^{\phantom{c}\nu\rho} T^c_{\phantom{c}\nu\lambda}
- \frac{h_a^{\rho}}{h} \mathcal{L}_G
+ \frac{1}{k} A^c_{\phantom{c}a\sigma} S_c^{\phantom{c}\rho\sigma}
\end{equation} 

A conservation law exists for $\Theta_a^{\phantom{a}\rho}$ and $\jmath_a^{\phantom{a}\rho}$
\begin{equation}\label{eq_conserv_law}
%Energy-momentum conservation law
\partial_{\rho}\left(kh \jmath_a^{\rho} + k\Theta_a^{\phantom{a}\rho}\right) = 0,
\end{equation}
that can be interpreted as a conservation law for the matter and the gravitational potential together. The gauge current $\jmath_a^{\rho}$ has a pseudotensor term $\frac{1}{k} A^c_{\phantom{c}a\sigma} S_c^{\phantom{c}\rho\sigma}$ which accounts for Lorentz non-inertial effects and can be eliminated by choosing a Lorentz inertial reference frame $\LorentzConn{a}{b}{\mu} = 0$. In (Lorentz) inertial frames a strict energy conservation law exists for matter and gravitation.

The Lagrangian density of matter is written using the Levi-Civita connection $\mathring{\omega}^a_{\phantom{a}b\mu}$ under the minimal coupling prescription $\mathcal{L}_M\left( \phi^i,\mathring{\nabla}_{\nu}\phi^i, \tetrad{a}{\mu} \right)$. Due to the independence of $\mathring{\omega}^a_{\phantom{a}b\mu}$ of $\LorentzConn{a}{b}{\mu}$, the Lagrangian density of matter depends only on the field, the tetrad, and their derivatives $\mathcal{L}_M\left(\phi^i,\partial_{\nu}\phi^i, \tetrad{a}{\mu}, \partial_{\nu}\tetrad{a}{\mu}\right)$. The total Lagrangian $\mathcal{L}_G + \mathcal{L}_M$ depends on $\LorentzConn{a}{b}{\mu}$ only in the total divergence of Eq.~\ref{eq_lagrangian_diverg} and consequently the Euler-Lagrange equations  $\delta \left( \mathcal{L}_G+\mathcal{L}_M\right)/\delta\LorentzConn{a}{b}{\mu}=0$ are trivially zero. However, the flattness of the spin connection is forced adding a Lagrange multiplier 
\begin{equation}
\mathcal{L}_R=\frac{h}{2\kappa} \lambda_a^{\phantom{a}b\mu\nu} R^a_{\phantom{a}b\mu\nu}
\end{equation}
which includes a kinetic term for $\LorentzConn{a}{b}{\mu}$. The resulting Euler-Lagrange equations only set constraints for the the Lagrange multipliers $\lambda_a^{\phantom{a}b\mu\nu}$ and leave the flat spin connection undetermined. 
\begin{equation}
\mathcal{D}_{\nu} \left(h\lambda_a^{\phantom{a}b\nu\mu}\right) = \kappa\Phi_a^{\phantom{a}b\mu} = 0
\end{equation}
The spin angular momentum of matter fields $\Phi_a^{\phantom{a}b\mu} = \delta \mathcal{L}_M / \delta \LorentzConn{a}{b}{\mu}$ is identically zero, what leaves $\lambda_a^{\phantom{a}b\mu\nu}$ decoupled from the matter fields. However, the terms coming from the variation of the Levi-Civita connection
\begin{equation}
\frac{\partial\mathcal{L}_{M}}{\partial \mathring{\omega}^c_{\phantom{c}d\sigma}} \frac{\partial \mathring{\omega}^c_{\phantom{c}d\sigma}}{\partial\left(\partial_{\nu}\tetrad{a}{\mu}\right)}
\end{equation}
are characteristic of Teleparallel theories in that they contribute to $\Theta_a^{\phantom{a}\mu}$.

\section{Transformations of the connection}\label{sec_conn}
%\section{Flat spin connections}~\label{sec_conn}
The set of all flat spin connections $\LorentzConn{a}{b}{\mu}$ form a subspace $D_{A}$ in the affine space of the metric spin connections. One point in $D_A$ can be parametrized with the 6 degrees of freedom of a local Lorentz transformation
\begin{equation}
A^{\phantom{0}a}_{0\phantom{a}b\mu} = \Lambda^{\phantom{c}a}_{c}\partial_{\mu}\Lambda^c_{\phantom{c}b}.
\end{equation}

On the other hand, the free components are counted after subtracting the null curvature conditions $R^a_{\phantom{a}b\mu\nu}=0$ from the 24 non-zero components of $\LorentzConn{a}{b}{\mu}$. The curvature tensor has 36 non-zero components and the second Bianchi identity $\partial_{\sigma}R^a_{\phantom{a}b\mu\nu} + \partial_{\mu}R^a_{\phantom{a}b\nu\sigma} + \partial_{\nu}R^a_{\phantom{a}b\sigma\mu} = 0$ sets 24 constraints, resulting only 12 independent. Therefore $\LorentzConn{a}{b}{\mu}$ is left with 12 free components, but only six degrees of freedom are required for local Lorentz transformations of $\LorentzConn{a}{b}{\mu}$. The other 6 degrees of freedom leave room to define an origin $A^{\phantom{0}a}_{0\phantom{a}b\mu}$ in $D_A$. A general $\LorentzConn{a}{b}{\mu}$ can be written as
\begin{equation}
\LorentzConn{a}{b}{\mu} = 
\InvLorTrans{b}{c}(x) \LorTrans{a}{d}(x) A^{\phantom{0}d}_{0\phantom{d}c\mu} + \InvLorTrans{c}{a}(x) \partial_{\mu} \LorTrans{c}{b}(x). 
\end{equation}

The Lorentz transformation will also act in all Minkowski tensors, including the tetrad $\tetrad{a}{\mu}$, but the capacity to define an origin in $D_A$ reflects the decoupling of the dynamics of the gravitational system from $\LorentzConn{a}{b}{\mu}$. In fact, a gauge transformation can be defined for $\LorentzConn{a}{b}{\mu}$ based in this property. For that purpose, it will be shown that a new flat Lorentz connection $\tilde{A}^a_{\phantom{a}b\mu}=\LorentzConn{a}{b}{\mu}+\mathcal{G}^a_{\phantom{a}b\mu}$ and Weitzenb\"ock spacetime connection $\tilde{\Gamma}^{\alpha}_{\beta\mu}=\SpaceTimeConn{\alpha}{\beta}{\mu}+\tetrad{b}{\beta}\Invtetrad{a}{\alpha}\mathcal{G}^{a}_{\phantom{a}b\mu}$ exist for the tensor
\begin{equation}
\mathcal{G}^{ab}_{\phantom{ab}\mu} = \InvLorTrans{c}{a}(x) \mathcal{D}_{\mu} \LorTrans{c}{b}(x), 
\end{equation}
with the covariant derivative defined using $\LorentzConn{a}{b}{\mu}$. The Lagrangian density of matter $\mathcal{L}_M$ is invariant under any gauge transformation of $\LorentzConn{a}{b}{\mu}$ and the Lagrangian density of TEGR $\mathcal{L}_G$ is also invariant up to a surface term which do not affect the equations of motion. Finally, the null curvature condition $\mathcal{L}_R$ remains trivially invariant because the new $\tilde{\Gamma}^{\alpha}_{\beta\mu}$ is also a flat connection. 

However, the kinematical transformation affects in a non-trivial way the torsion. 
\begin{equation}
\tilde{T}^a_{\phantom{a}\mu\nu} = T^a_{\phantom{a}\mu\nu} 
+ \mathcal{G}^{a}_{\phantom{a}b\mu}\tetrad{b}{\nu}
- \mathcal{G}^{a}_{\phantom{a}b\nu}\tetrad{b}{\mu}
\end{equation}
To analyze how the torsion changes, the tetrad will be decomposed in a trivial tetrad fullfilling $\FockIvanenko{\nu}\Trivtetrad{a}{\mu}=0$ and an anholonomic term $\Bfield{a}{\mu}$. 
\begin{equation}~\label{eq_tetrad_decomp}
\tetrad{a}{\mu}=\Trivtetrad{a}{\mu}+\Bfield{a}{\mu}
\end{equation}
Under this decomposition the torsion is written as
\begin{equation}\label{eq_Bfield_define}
\Torsion{a}{\mu}{\nu} = 
%\mathcal{D}_{\mu} \tetrad{a}{\nu}(x) - \mathcal{D}_{\nu} \tetrad{a}{\mu}(x) = 
\mathcal{D}_{\mu} \Bfield{a}{\nu}(x) - \mathcal{D}_{\nu} \Bfield{a}{\mu}(x).
\end{equation}
After a transformation of $\LorentzConn{a}{b}{\mu}$, the trivial tetrad and the anholonomic field acquires a global acceleration field $\alpha^a_{\phantom{a}\mu}$ 
\begin{subequations}
\begin{eqnarray}
\tilde{B}^a_{\phantom{a}\mu}=\Bfield{a}{\mu} + \alpha^a_{\phantom{a}\mu}
\\
\tilde{e}^a_{\phantom{a}\mu} = \Trivtetrad{a}{\mu} - \alpha^a_{\phantom{a}\mu}
\end{eqnarray}
\end{subequations}
determined by the conditions 
\begin{subequations}
\begin{eqnarray}
\tilde{\mathcal{D}}_{\nu} \tilde{e}^{a}_{\phantom{a}\mu} = 0
\\
\tilde{h}^a_{\phantom{a}\mu}=\tetrad{a}{\mu}
\end{eqnarray}
\end{subequations}
The condition determining $\alpha^a_{\mu}$ given $\mathcal{G}^a_{\phantom{a}b\nu}$ do not depend on the tetrad or the gravitational potential $\Bfield{a}{\mu}$
\begin{equation}\label{eq_B_A}
\mathcal{D}_{\nu}\alpha^a_{\phantom{a}\mu} = \mathcal{G}^a_{\phantom{a}b\nu} \left(\Trivtetrad{b}{\mu}-\alpha^b_{\phantom{b}\mu}\right)
\end{equation}
keeping the relation valid under coupling to matter fields. The new torsion $\tilde{T}^a_{\phantom{a}\mu\nu}$ is
\begin{eqnarray}
\tilde{T}^{a}_{\phantom{a}\mu\nu} = \frac{1}{2} T^a_{\phantom{a}\mu\nu} 
+ \LorentzConn{a}{b}{\mu}\alpha^b_{\phantom{b}\nu}
+ \mathcal{G}^{a}_{\phantom{a}b\mu}\left(\Bfield{b}{\nu}+\alpha^b_{\phantom{b}\nu}\right) - \left( \mu \leftrightarrow \nu \right).
\end{eqnarray}
The tensor $\mathcal{G}^a_{\phantom{a}b\mu}$ generates a gauge transformation of $\LorentzConn{a}{b}{\mu}$ which introduces a global acceleration field in the reference frame which does not change the dynamics of the gravitational field and the matter fields. Moreover, the gauge symmetry allows, at least at classical level, choosing a zero Lorentz connection and cancel the inertial effects. However, $\LorentzConn{a}{b}{\mu}$ remains as part of the TG, because there is not other connection within the theory which could replace it.

\section{Discussion}~\label{sec_discuss}
The most general gauge transformation of $\Bfield{a}{\mu}$ in the internal space which leaves the Lagrangian, and the equation of motions, of TEGR invariant is $\delta\Bfield{a}{\mu} = \alpha^{a}_{\phantom{a}\mu} + \mathcal{D}_{\mu}\epsilon^a$. While the first term comes from transformations of $\LorentzConn{a}{b}{\mu}$, the second term is a zero torsion term that depends in the vector $\epsilon^a$ and can be related to a local translation in the internal space~\cite{aldrovandi_teleparallel}. While the translations in the internal space leave the torsion invariant, the transformation of $\LorentzConn{a}{b}{\mu}$ is a kinematical symmetry valid at Lagrangian level, similarly to the global Lorentz transformations in SR. 

It is worth remarking that both are gauge symmetries of the internal Minkowski space and as thus, they are missing in GR. There, the univocal definition of the Levi-Civita connection from the metric (or, equivalently, the tetrad) prevents the existence of inertial frames in a general case. A consequence of this, is the necessity of defining covariant conservation laws for the Energy-Momentum tensor, instead of the strict conservation law of equation~(\ref{eq_conserv_law}). Although, a non-zero $\LorentzConn{a}{b}{\mu}$ will result in a diverging asymptotic value for the energy-momentum tensor, the gauge symmetry of the flat spin connection ensures that for any value of $\tetrad{a}{\mu}$, it is always possible to set to zero $\LorentzConn{a}{b}{\mu}$ and leave the conservation law for the energy-momentum free of inertial effects. The implications for a quantum theory of a kinematical connection that is not observable, at least not by the gravitational effects exerted on matter fields, are not clear. The future study of the canonical form of TEGR including the $\LorentzConn{a}{b}{\mu}$ symmetry will clarify this aspect. 

After a $f(R)$ modification the transformation of $\LorentzConn{a}{b}{\mu}$ remains a kinematical symmetry of the full Lagrangian. Contrary, a general $f(T)$ modification of the TEGR Lagrangian breaks the $\LorentzConn{a}{b}{\mu}$ symmetry while keeping the matter Lagrangian invariant. Implying that non-equivalent torsion tensors exert the same gravitational effect on matter and a non-zero $\LorentzConn{a}{b}{\mu}$ is not observable directly by measuring its influence on matter. A similar behaviour of the torsion in the experimentally viable one parameter family of Teleparallel theories was considered unphysical by Kopczynski~\cite{kopczynski_problems_1982} and Nester~\cite{nester_is_1988}. On the other hand, a broken symmetry in the gravitational vacuum can be a source of rich phenomenology with cosmological implications which could shed light into the origin of dark matter or the dark energy. This also justifies further studies that deepens on the effect of a broken symmetry in the gravitation vacuum for a non-observable connections. 

\section{Conclusions}\label{sec_conc}
The necessity of non-zero $\LorentzConn{a}{b}{\mu}$ in TEGR has been discussed. It has been shown that its variations leaves the total Lagrangian unchanged. The Euler-Lagrange equations for $\LorentzConn{a}{b}{\mu}$ leave undetermined the flat spin connection, even in the presence of matter fields. Moreover, in the minimal coupling prescription of the gravitational field to the matter fields, compatible with the equivalence principle, the Levi-Civita connection needs to be used and since, it does not independent on the flat spin connection, the spin angular momentum of matter fields is shown to be identically zero within TEGR. Although, the energy-momentum tensor is shown to have extra terms coming from the variation of the Levi-Civita connection respect to the tetrad. 

A kinematical gauge symmetry of the flat spin connection which modifies the torsion but leaves the Lagrangian density of TEGR invariant has been identified. The changes of the torsion have been analyzed defining an anholonomic field $\Bfield{a}{\mu}$ and it has been shown that gauge transformations of the flat spin connection, only introduces global acceleration fields which can be interpreted as manifestation of a local Lorentz transformation. The gauge symmetry makes the conservation law of the energy-momentum of the gravitational field and matter of TEGR free of non-inertial effects. At the classical level, the dynamical equations can be studied making zero trivially the flat spin connections, but considering a non-observable flat spin connection can be necessary to build a consistent quantum theory of gravitation. Any $f(R)$ modification of TEGR will maintain the gauge symmetry but a general $f(T)$ modification will break it for the Lagrangian of the gravitational field while keeping the Lagrangian density of matter invariant. Further studies may decide if a broken symmetry of the vacuum of gravity is unphysical or if it can be a source of new phenomenology.

\end{document}